\documentclass[sigconf, nonacm=true, balance=true]{acmart}

\usepackage[utf8]{inputenc}
\usepackage{import}
\usepackage{csquotes}
\usepackage{balance}
\usepackage{graphicx}
\usepackage{booktabs}
\usepackage{hyperref}

\hypersetup{
pdftitle={Where Do We Meet? Key Factors Influencing Collaboration Across Meeting Spaces},
pdfsubject={Hybrid Meetings},
pdfauthor={Isaac Valadez, Sandra Trullemans and Beat Signer},
pdfkeywords={Collaboration, In-Person Meetings, Virtual Meetings, Hybrid Meetings, Cross-Space Collaboration Model}
}

\usepackage{balance}

\AtBeginDocument{%
  \providecommand\BibTeX{{%
    \normalfont B\kern-0.5em{\scshape i\kern-0.25em b}\kern-0.8em\TeX}}}

 \newcommand{\dquote}[1]{``#1''}

 \newcommand{\feature}[1]{\noindent \textbf{#1} \\}
 
 \newcommand{\myquote}[1]{\emph{\dquote{#1}}}
 \newcommand{\head}[1]{\vspace{1.5mm} \noindent \textbf{#1}\\}

\setcopyright{acmcopyright}
\copyrightyear{2018}
\acmYear{2018}
\acmDOI{XXXXXXX.XXXXXXX}

\acmConference[DIS 2023]{Make sure to enter the correct
  conference title from your rights confirmation emai}{July 10--14,
  2023}{Pittsburgh, USA}
\acmBooktitle{DIS '23: ACM Designing Interactive Systems,
 July 10--14, 2023, Pittsburgh, USA} 
\acmPrice{15.00}
\acmISBN{978-1-4503-XXXX-X/18/06}

\begin{document}

\title{Where Do We Meet? Key Factors Influencing Collaboration Across Meeting Spaces}


\author{Isaac Valadez}
\affiliation{
  \institution{Innoty}
  \streetaddress{Nachtegaallaan 4}
  \city{Zellik}
  \country{Belgium}
  \postcode{1731}
}

\affiliation{%
  \vspace{0.15cm}
  \institution{WISE Lab, Vrije Universiteit Brussel}
  \streetaddress{Pleinlaan 2}
  \city{Brussels}
  \postcode{1050}
  \country{Belgium}
}
\email{jvaladez@vub.be}
\orcid{0000-0003-2668-5768}

\author{Sandra Trullemans}
\affiliation{
  \institution{Innoty}
  \streetaddress{Nachtegaallaan 4}
  \city{Zellik}
  \country{Belgium}
  \postcode{1731}
}
\email{sandra@innoty.biz}

\author{Beat Signer}
\affiliation{%
  \institution{WISE Lab, Vrije Universiteit Brussel}
  \streetaddress{Pleinlaan 2}
  \city{Brussels}
  \postcode{1050}
  \country{Belgium}
}
\email{bsigner@vub.be}
\orcid{0000-0001-9916-0837}


\begin{abstract}
Over the past years, there has been a shift towards online and hybrid meeting forms in workplace environments, partly as a consequence of various COVID-19 restrictions. However, the decision-making process on how to best collaborate with team members is predominantly driven by practical concerns. While there is a significant body of literature about where to best meet, this knowledge is fragmented across various disciplines and hard to use in novel meeting solutions. We present the Cross-Space Collaboration model which identifies the main factors that drive the features of in-person collaboration and the meeting aspects that influence these factors such as cognitive load. We designed the model to give guidance to teams and individuals on how to meet in order to have a higher collaboration effectiveness. Finally, we outline how the model can bring added value within new meeting solutions, next generation virtual reality meeting spaces and educational settings.
\end{abstract}




\maketitle


\section{Introduction}

During the COVID-19 pandemic, many companies were pushed towards remote and hybrid work. Although the originally strong restrictions have been lifted in many countries, various companies have decided to keep some form of hybrid work policy or, in some cases, switch to a complete remote scheme. Perhaps more important is the fact that a high percentage of workers now expect or request flexible working policies from their employers. This transition introduces a number of challenges for teams. While remote collaboration offers multiple benefits in terms of convenience, important benefits from physical collaboration are often missing. This includes both, direct benefits such as participants' improved ability to understand each other's emotions, as well as indirect effects, including post-meeting informal discussions opening the way for new ideas and increased team bonding. 

We can observe that teams and individuals make a cost-benefit analysis every time they consider whether to meet physically. In these cases, they seem to consider personal preferences, practical concerns and company policies. However, we might ask ourselves whether they take into account if the chosen meeting space aligns with their goals for collaboration; are they further aware of the cognitive implications of collaborating in physical or digital space? Are they aware of the short- and long-term benefits of meeting physically rather than digitally?

In order to bring some light into the relationship between collaboration spaces, goals and participants, in this paper we are going to address the following research questions:

\begin{itemize}
\item RQ1: What are the main factors in physical collaboration that benefit knowledge transfer effectiveness? 
\item RQ2: Which aspects of a meeting influence these main factors?
\item RQ3: How can the presented Cross-Space Collaboration model inform the design of future tools for meetings and collaboration?
\end{itemize}

After introducing some related work in the domain of collaboration, cognitive science and embodied cognition, we present the Cross-Space Collaboration model with its three main dimensions and discuss the key factors of physical collaboration influencing knowledge transfer across spaces. Further, we briefly discuss a preliminary evaluation where industry professionals have been interviewed to find potentially missing key elements in our model. After presenting some extensions of the model, we highlight a number of possible use cases for the Cross-Space Collaboration model. Finally, we discuss some current limitations and outline future work to further validate the model.

\section{Background}

The COVID-19 pandemic has greatly influenced research on the impact of working from home, in the office and in hybrid environments. New studies have been conducted to better understand the impact of video conferencing~\cite{zoom_fatigue,educational_videoconferencing}, working remotely~\cite{yang_effects_2022,forced_flex,chatterjee_does_2022} and collaboration via hybrid work~\cite{neumayr_what_2021,microsoft_2021}. Additionally, we have seen efforts to bring these insights together~\cite{teevan2022microsoft,neumayr_what_2021}. 

At the same time, we have seen a surge of investment in meeting and collaboration technology as companies and academia look for new ways to facilitate remote work and virtual collaboration. On one side, there have been major efforts to improve remote collaboration technologies through traditional means, namely improving noise reduction, latency and other technical aspects that can hinder collaboration. On the other hand, we have seen the addition and improvement of features such as integrated virtual whiteboards and different layouts to view meeting participants. There has even been a push towards the research and development of immersive collaborative experiences using virtual reality~\cite{olaosebikan_embodied_2022,steinicke_pilot} and augmented reality~\cite{chen_effect_2021,model_collaborative_AR}. 

Despite these studies and advancements, we still see teams and companies struggling to find the optimal balance between remote, co-located and hybrid work. We believe that this is partly caused by the disconnection between findings and the complexity of understanding how they relate to each other. By making visible what is at the core of these findings, we can help teams to enhance the way they collaborate. We can further inform designers and developers on the most important factors to consider when developing tools for collaboration and knowledge transfer. 

In the past, different tools and methods have been created to bridge theory and practice. Notions such as \emph{bridging concepts} which bring together theory, design implications and practical examples~\cite{bridging_concepts}; toolkits which translate theory into easy-to-digest formats and position the theory in specific contexts to inform designers~\cite{tac_toolkit,tango}; and frameworks that identify common themes, propose definitions and unify understanding in a variety of topics, including tangible interaction~\cite{grasp_framework}, engagement~\cite{engagement_gamification} and collaboration~\cite{patel_factors_2012}.

In particular in the domain of collaboration, there are multiple relevant frameworks. For instance, the CoSpaces Collaborative Working Model~(CCWM) considers key factors including \emph{interaction processes}, \emph{context} and \emph{individuals}~\cite{patel_factors_2012}. The Theory of Remote Scientific Collaboration~(TORSC)~\cite{remote_scentific} looks at factors that influence successful remote scientific collaboration, such as common ground or collaboration readiness, while the Domino framework describes groups of collaborators and their dependency between each other throughout various moments of collaboration~\cite{domino}. 

What is missing is a new approach that looks at collaboration from a meeting-centred perspective that considers the complete context of how the meeting happens (i.e.~interaction dynamics), where it happens (i.e.~space), who is participating and what the meeting participants are trying to achieve. This is in line with Saatci~et~al.'s~\cite{saatci_reconfiguring_2020} previous proposal to move from an actor-activity approach to a phenomenon-centred approach looking at meetings as something that can be configured and designed for success.

\section{Cross-Space Collaboration Model}
Knowledge transfer is a broad concept that can cover multiple levels, such as people-to-people, people-to-organisation and organisation-to-organisation knowledge transfer. The Cross-Space Collaboration model focuses on identifying factors of people-to-people in-meeting collaboration that make physical (co-located) collaboration more beneficial than digital collaboration in terms of knowledge transfer. Additionally, it describes which meeting aspects have an influence on these factors. 

As a result, our model can for instance be used to design a meeting recommendation solution for providing users insights during their cost-benefit analysis when scheduling a meeting. In this scenario, many individuals perform a cross-benefit analysis, even if informally, to decide where to meet with their team. Such an analysis ideally consists of the following steps: considering the purpose of the meeting, identifying the alternatives (e.g.~space and tools), deciding whose benefits and costs count, identifying and predicting the impact of the decision, computing the value of each alternative and, finally, taking a decision~\cite{boardman}. Our model can be used to easily compute the benefits and costs of the meeting alternatives and provide users with insights to make a decision. Additionally, the model might be used to provide guidance on how to moderate the meeting by using different tools.           

\begin{figure*}[htb]
    \centering
    \includegraphics[width=1.0\textwidth]{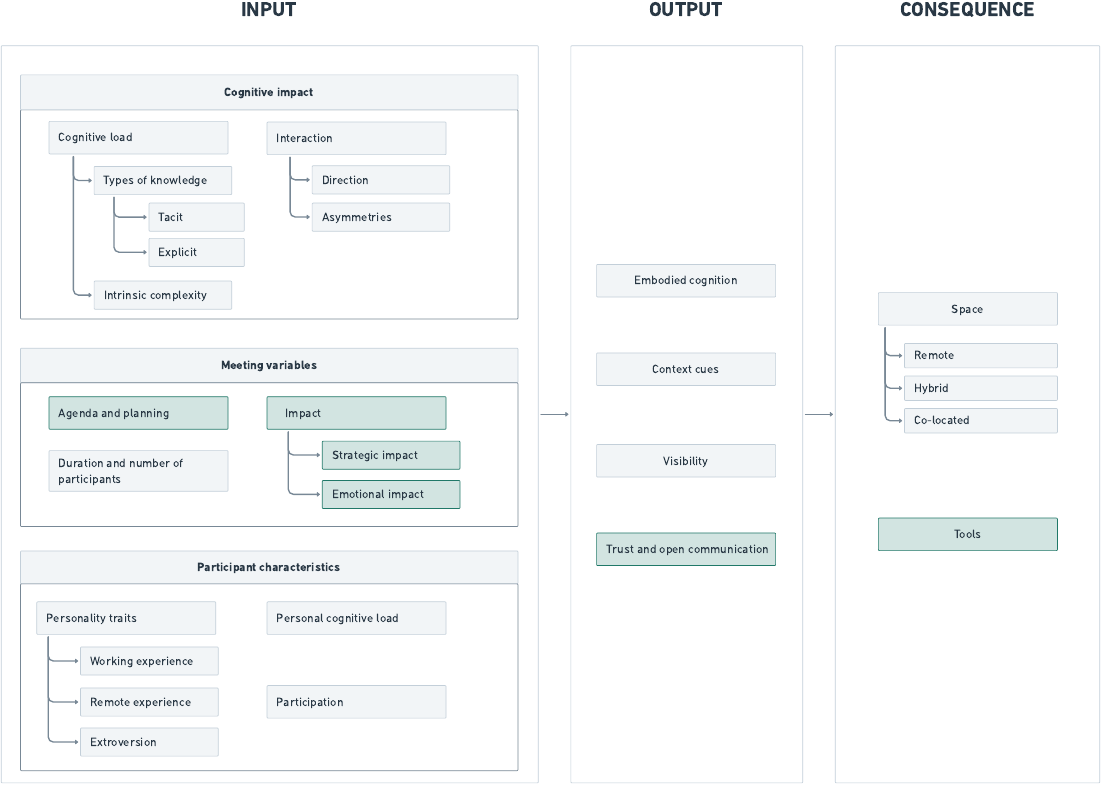}
	\caption{Cross-Space Collaboration model with meeting aspects (input) that influence beneficial factors of in-person collaboration (output) and how these translate into concrete actions (consequences). Elements in green were added to the model after a preliminary evaluation.}
  \label{fig:crossSpaceModel}
\end{figure*}

Our Cross-Space Collaboration model highlighted in Figure~\ref{fig:crossSpaceModel} consists of an output side that covers the main beneficial factors for in-person collaboration and an input side representing the meeting aspects that influence these main factors on the output side. The higher the value of the main factors on the output side, the more beneficial the physical space and tools offered by in-person collaboration become. While collaboration is established by multiple persons (i.e.~on a team level), the model does take the individual into account. This implies that the main factors on the output side can be valued for a team as well as specific team members. By having these two perspectives, the model can be applied to guide the user in managing a hybrid meeting where certain team members better attend the meeting in person while others might join online.    

\subsection{Modelling Approach}
In order to identify the main factors and their influencing meeting aspects, we have analysed a major body of related work. We did a keyword search in the ACM Digital Library and on Google Scholar. We started our search with key concepts such as \dquote{knowledge transfer}, \dquote{knowledge processing}, \dquote{collaboration}, \dquote{remote collaboration}, \dquote{hybrid collaboration}, \dquote{multimodal learning}. As we analysed the initial literature, we increased our scope to include other keywords such as \dquote{embodied cognition}, \dquote{data visualisation}, \dquote{data physicalisation}, \dquote{knowledge management}, \dquote{non verbal communication} and \dquote{future of work}. We then conducted a semantic search to increase our understanding of the topics and to open our perspective on key concepts. 

From our first analysis, we identified the core dimensions for each side of the model. Subsequently, the findings were classified according to their relevance to a dimension as well as their influence on the main factors. In this step, dimensions were also sub-categorised when they individually had a significant impact on the main factors. In the following, we provide a discussion of all the identified main factors and their influencing meeting aspects. 

Note that the completeness of the Cross-Space Collaboration model has been analysed in a preliminary exploratory user study. The study results led to some minor extensions of the initial model with the four additional factors shown in green in Figure~\ref{fig:crossSpaceModel} and discussed later in Section~\ref{sec:adaptation}.

\subsection{Main Features of Physical Collaboration}
In this section, we introduce the three main factors that make in-person collaboration more beneficial than digital collaboration with regards to its knowledge transfer effectiveness. Thereby, we provide an answer to the earlier formulated research question~RQ1. Note that the higher the value of these three factors shown in the middle of Figure~\ref{fig:crossSpaceModel}, the more a team or individual benefits from meeting in person.   

\head{Embodied cognition} Besides semantic and episodic memory, we use embodied cognition while processing information. Embodied cognition refers to the perceptual and motor systems providing memory cues used at the time of receiving information and during recall~\cite{madan_using_2012,wakefieldLearning2019}. The manipulation of physical objects allows us to be more expressive in transferring the message and decreases verbal effort. It also helps us to follow our line of thought~\cite{hornecker_understanding_nodate}. In virtual space, embodied cognition is often limited to facial expressions. Furthermore, studies have shown that physical items have a higher impact on recall than digital ones as they tend to be more prominent and noticeable~\cite{things_that_make_us_reminisce,van_de_hoven}. However, in virtual reality solutions, there is a major effort in developing avatars that can express a person's gestures and emotions in real time~\cite{vreed,ahuja_moves,controllerpose_ahuja,namikawa2021emojicam}. Another effort is to provide realistic artefacts that stimulate embodied cognition \cite{recompfig,Bozgeyikli_tangiball}. Note that it is important to take the use of embodied cognition into account when the cognitive effort is high during collaboration.         

\head{Context cues} In synergy with embodied cognition, episodic memory plays a significant role during recall. The lifelogging paradigm is one of the well-studied approaches to support episodic memory~\cite{things_that_make_us_reminisce,dingler_memory_2021}. Besides time-based recall, spatial cues are also heavily used and researched within the field of personal information management~\cite{spatial_jones,haystack}. However, studies in various fields such as notetaking \cite{prosthetic_memory} and the re-finding of documents~\cite{kalnikaite_beyond_2010,prosthetic_memory} have indicated that the contextual cues have the most impact on recall. The physical environment provides enough contextual triggers to allow us to recall details after 30~days while the digital space limits this recall to 7~days. This is one of the reasons why working fully remotely implies more follow-up meetings~\cite{microsoft_2021}. Within meetings, the context cue is triggered by events in our surroundings such as team behaviour, heated moments in discussions or the movement of artefacts. Non-verbal cues, such as gestures and body language, also impact communication by signalling emotions, agreement and disagreement~\cite{microsoft_2021}. We can conclude that when recall is important in the collaboration process, the physical environment provides more contextual triggers than the digital one.

\head{Visibility} The ability to visualise information from multiple sources in different formats and to have a shared view of said information has a positive impact on a team's ability to solve problems, overcome cognitive restrains, handle cognitive load and recall information~\cite{Knowledge_vis_sahli,recall_tech_mediated, eppler_visualization_2013}. This impact comes from the visual material itself but also from the ability of other team members to see how knowledge is being processed and manipulated. We can also observe the advantages of visibility in tangible collaboration toolkits. Over the past years, research has been investigating how so-called physical ideation toolkits can enhance creativity, sense-making and discovery, making use of cards\cite{tango,tac_toolkit,iot_cards}, Playmobil~\cite{playmobil} or Lego~\cite{kristiansen2014building}. The use of Playmobil and other 3D~printed artefacts has also been proposed to guide daily in-person meetings where a shared whiteboard is augmented with these artefacts to keep a unified and visible overview of the meeting's content during discussions involving various profiles~\cite{playmobil}. While in common digital meeting solutions visibility is limited to a shared screen or digital whiteboard, efforts are being made to provide more visual cues of, for example, the speakers~\cite{murali_affectivespotlight_2021}. However, in collaborative augmented reality solutions, a unified view of the meeting's material/content is one of the major design focus points to make collaboration more effective~\cite{chen_effect_2021,model_collaborative_AR}. We can conclude that physical unified visibility aids participants when the collaborative activity involves creativity, sense-making, open mindset and discussions between different stakeholders and profiles.

\subsection{Influencing Meeting Aspects}
On the left-hand side of the model, we defined the meeting aspects that influence the value of the main factors outlined in the previous section. The meeting aspects are categorised in three dimensions, namely cognitive impact, meeting variables and participant characteristics. In the following, we outline each dimension and their impact on the value of the main factors beneficial to in-person collaboration as summarised in Table~\ref{tab:my-table}.

\subsubsection{Cognitive Impact}
Within this first dimension we have identified the two main meeting aspects of cognitive load and interaction. Cognitive load handles \emph{what} kind of information is being discussed during the collaborative activity while interaction focuses on \emph{how} the information is transmitted among the team members.

\head{Cognitive load} With cognitive load we refer to how the nature of information or knowledge has an effect on the difficulty of processing the information by the team members. The most relevant theory to the presented model is the work of Polanyi~et.al.~\cite{rovik_knowledge_2016} that describes two main knowledge types, explicit and tacit knowledge. The key discriminator between both types is the level at which they can be verbalised. Thereby, \emph{explicit knowledge} is easy to verbalise as it is often represented by structured and documented knowledge (e.g.~a how-to book). On the other hand, \emph{tacit knowledge} is hard to verbalise and unstructured (e.g.~opinions shared in a brainstorm meeting). Explicit knowledge thereby often imposes a lower cognitive load on the recipient of information than tacit knowledge. Additionally, the \emph{intrinsic complexity} of the knowledge defined by its structure, the material's complexity and the number of knowledge units that need to be linked simultaneously to be understood, also play a significant role in impacting the cognitive load~\cite{brunken_direct_2003,schmidt_modeling_2016}. The intrinsic complexity is, for example, high for meeting participants where an expert presents a new AI~solution to stakeholders with a diverse background. The solution may be easy to verbalise by the expert (i.e.~explicit knowledge) but other participants might experience a high intrinsic complexity when trying to understand the expert.    

\underline{Impact:} 
In order to decrease the cognitive load, all kinds of prosthetic memory aids such as sketching, notes or post-its have been used for decades. This act of using physical and digital aids is known as cognitive offloading~\cite{risko_cognitive_2016}. The Extended Mind framework~\cite{clark_extended} defines nine principles to manage cognitive offloading where among others it is important to offload knowledge into the real world (e.g.~whiteboard or notetaking) and to encode knowledge into artefacts and interact with them to enhance the understanding such as done in tangible design toolkits \cite{tac_toolkit} or educational serious play solutions~\cite{kristiansen2014building, playmobil} and even virtual reality notetaking tools~\cite{olaosebikan_embodied_2022}. We can conclude that the higher a meeting participant's cognitive load, the more they will profit from embodied cognition and visibility.

\head{Interaction} 
The interaction describes the way how people interact with each other to share knowledge and its impact on knowledge transfer. Interactions can be unidirectional or bidirectional, depending on whether both parties in a conversation act as senders and recipients of information~\cite{tangaraja_knowledge_2016}. Since collaboration is an act of two or more persons, we focus on bidirectional interactions. The work of Heath and Luff~\cite{disembodied} provides some insights on the asymmetries of interaction in media spaces (i.e.~hybrid and digital). These asymmetries are mainly driven by changes in the interplay and the way of communication in video-mediated meetings. Recent research by Saatci~et~al.~\cite{saatci_reconfiguring_2020}~builds on these research findings and identifies asymmetries around the technical, social and cultural interaction in hybrid settings. Among others, they observed that remote participants were isolated and co-located participants dominate the conversations. Further, cultural differences such as waiting to speak versus allowing for interruption were more visible. It is also known from existing work that within hybrid settings the primary room issue, where the physical space remains dominant in the interaction, plays an important role in disconnected interactions~\cite{karis_remote}. Moreover, in-person meetings have fast-moving conversations, people talk through each other, have implicit floor control (e.g.~one can take the word via language) and significant side conversations between people take place~\cite{karis_remote,isaacs_tang}. These interactions are important for collaborative tasks but they are lacking in digital and hybrid settings. Thereby, tasks like decision making, creative work, brainstorming or planning are less efficient in technology-mediated settings~\cite{teevan2022microsoft}.                 

\underline{Impact:} 
When an intensive bidirectional interaction will take place, the physical space is still the most opportune medium to meet. Additionally, we must not underestimate the power of the informal interactions that take place during and after physical meetings~\cite{microsoft_2021}. They are crucial for a company's long-term innovation roadmap and to establish the powerful weak ties network that drives informal discussions and innovation~\cite{teevan2022microsoft,granovetter}. Visibility of the meeting content by using tools such as the Sticx boards~\cite{sticx}, can help to keep track of the asynchronous discussions' outcomes within physical meetings. However, using physical or digital tools may contribute to the asymmetries of interactions within hybrid settings as shown by Saatci~et~al.~\cite{saatci_reconfiguring_2020, neumayr_what_2021} where the use of physical tools such as post-it notes disconnects remote participants and the use of virtual tools like Mural~\footnote{https://www.mural.co} eliminates the physical benefits for co-located persons. Choosing the right tool to support the interaction in hybrid or digital spaces also comes with cognitive and time costs as it has to be compared to other solutions, learned and set up~\cite{teevan2022microsoft}. Within the physical space these are less of a hurdle. We can conclude that for intensive bidirectional interactions, visibility and context cues need to be supported in co-located meetings. When a digital or hybrid meeting is implied by surrounding factors, we must be aware of the asymmetries of interaction.

\subsubsection{Meeting Attributes}
Within the meeting attributes dimension, we focus on how the meeting format and conditions impact the sharing of knowledge, regardless of the topic that is being discussed. 

\head{Duration and number of participants}
Studies by Microsoft during the COVID-19 pandemic~\cite{microsoft_2021} have shown that collaboration over video calls adds extra mental load and fatigue compared to co-located collaboration. The video call fatigue can be caused by the lack of new stimuli, failures in technology and delay in a call or even the continuous observation of ourselves on the screen. Additionally, the number of participants has a significant influence on the quality of interaction within digital settings. Already for small teams, the studies revealed challenges with speaking into the void and reading the room to get a grasp of the meeting dynamics. The social presence is also harder to achieve when the number of participants increases as their video feeds get squeezed into small tiles. Research by Saatci~et~al.~and others~\cite{saatci_reconfiguring_2020,neumayr_what_2021} further highlights the challenge of turn-taking. In physical space, a more natural interaction is set regardless of the number of participants within the team (i.e.~the physical round table or interruptions imposed by body language) while remote participants are more formally given the word by raising their digital hand or by calling their name. However, when a significant number of participants is reached (e.g.~more that 100~participants), the asymmetries in interaction decrease as interaction gets less attention and is switched towards chat functionalities~\cite{microsoft_2021}.  

\underline{Impact:}
For longer meetings, teams might consider to plan in-person meetings to reduce the mental load of video-mediated meetings. Additionally, it is important to provide participants with significant embodied cognition, context cues and visibility to compensate the cognitive load implied by long meetings. A common physical view (e.g.~whiteboard sketch or Playmobil scene~\cite{sticx}) on what has been discussed can aid the alignment of a team since for longer meetings this might be challenging due to, for instance, the side conversations or interruptions as the nature of interaction. Finally, when remote participants need to be involved, unifying turn-taking solutions need to be considered such as claiming the speech by \dquote{raising the hand} on a smartphone app regardless of the space where the person is~\cite{saatci_reconfiguring_2020}.

\subsubsection{Participant Characteristics}
The participant characteristics are factors that focus on the individuals participating in the meeting. They concern personal traits, their individual cognitive load and level of participation. 

\head{Personal traits}
Major research has been conducted over the past decades on aspects of personal traits during meetings. We thereby only take the main findings into account that might influence the main outcome factors. 

A first aspect to consider is the \emph{working experience} of the participant. For junior team members, it tends to be difficult to find an opportunity to ask questions or participate in decision-making processes. This is even more pronounced in digital environments~\cite{sarkar_promise_2021}. Team members might further be less familiar with certain meeting rituals or methodologies that are in place by the team~\cite{microsoft_2021}. A second aspect is the \emph{remote experience} of participants. While during/after the pandemic technology literacy has decreased significantly, there are still technical asymmetries of interaction that often need to be overcome~\cite{teevan2022microsoft,saatci_reconfiguring_2020,neumayr_what_2021}. Technology might fail, connection issues occur or video feeds might freeze. While these are only the daily issues that need to be handled, the introduction of advanced collaboration solutions imposes another level of experience for remote participants. Figuring out how to use the solutions and getting them ready might lead to mental stress and dissatisfaction even before a meeting starts~\cite{saatci_reconfiguring_2020}. Finally, the participant's \emph{level of extroversion} might be considered in deciding to follow a meeting in person or not. Moreover, participants who are highly extroverted can take over discussions and block the participation of more introverted attendees. Within digital tools, features such as \dquote{raising the hand} can give those participants a chance to be part of the discussion~\cite{sarkar_promise_2021,das_towards_2021}. On the other hand, non-verbal cues such as leaning forward or clearing one's throat may go unnoticed during virtual meetings. However, research from Microsoft indicates that more introverted personalities do not significantly use the \dquote{raising the hand} feature but rather start a conversation with trusted persons in a chat as a side channel to the meeting \cite{microsoft_2021}. 
 
\underline{Impact:}
Individuals with limited work experience might benefit from greater visibility into the meeting's content and attendees; this increases their chance to communicate their contributions. Visibility may lower the barrier as participants can point to or interact with an artefact in a group context. In addition, embodied cognition and the provision of context cues facilitate knowledge processing and are crucial for new team members as discussed earlier. Within hybrid meetings, participants with lower extroversion and remote experience may benefit from attending in person for highly collaborative meetings as the physical space implies more informal communication and natural floor control~\cite{isaacs_tang}.

\feature{Personal cognitive load}
Learning a new topic imposes cognitive load on people as they need to process new information. This load depends on the intrinsic complexity of the topic but is also influenced by the familiarity an individual has with the topic or other related topics~\cite{johnson_exploring_2010}. In general, people learn better through interactive learning~\cite{chi_translating_2021} and, as mentioned before, benefit from embodied cognition. 

\underline{Impact:}
When people are going to be introduced to new topics and in particular if the complexity of the topic is high, physical spaces seem to be better at fostering interactive learning. In contrast, for topics attendees are familiar with and where there is therefore a lower cognitive load, virtual meetings can work as well.

\begin{table*}[t]
\normalsize
\resizebox{\textwidth}{!}{%
\begin{tabular}{|ll|}
\hline
\multicolumn{1}{|l|}{\textbf{Aspect}}                     & \textbf{Impact}                                                                                                                                                                                                                                                                                                                                                                                                                                                                                                                                                               \\ \hline
\textit{Cognitive Impact}                                 &                                                                                                                                                                                                                                                                                                                                                                                                                                                                                                                                                                               \\ \hline
\multicolumn{1}{|l|}{Cognitive load}                      & - Meetings which impose a high cognitive load on participants benefit from embodied cognition and visibility.
\\ \hline
\multicolumn{1}{|l|}{Interaction}                         & \begin{tabular}[c]{@{}l@{}}- Meetings with intensive bidirectional interactions benefit from visibility and context cues. \\ - Meetings with unidirectional interaction are less dependent on embodied cognition and context cues.\end{tabular}                                                                                                                                                                                                                                                                                                                        \\ \hline
\textit{Meeting variables}                                &                                                                                                                                                                                                                                                                                                                                                                                                                                                                                                                                                                               \\ \hline
\multicolumn{1}{|l|}{Agenda and planning}                 & \begin{tabular}[c]{@{}l@{}}- Meetings with agendas that are not highly structured benefit from embodied cognition and flexible tools.\\ - Meetings with more flexible agendas can take advantage of embodied cognition to ease improvisation.\end{tabular}                                                                                                                                                                                                                                                                                                                  \\ \hline
\multicolumn{1}{|l|}{Duration and number of participants} & \begin{tabular}[c]{@{}l@{}} - Long digital meetings impose a higher cognitive load on participants than their physical counterparts. \\ - Communication challenges increase as the number of participants rises. These challenges are eased by \\ \hspace{0.17cm}visibility and embodied cognition (until the number of participants gets too large).
\end{tabular}                                                                                                                                                                                                                              \\ \hline
\multicolumn{1}{|l|}{Strategic and emotional impact}      & \begin{tabular}[c]{@{}l@{}}- Meetings with high strategic impact require open communication and knowledge share which profits form \\ \hspace{0.17cm}embodied cognition. Visibility of information is key to have a shared understanding of the problem or project.\\ - Context cues will help team members to remember agreements and their context for an increased period \\ \hspace{0.17cm}of time.\\ - Strong emotions increase cognitive load. \\ - Participants may have personal preferences on how and where to handle topics that concern privacy \\ \hspace{0.17cm}or are considered personal.\end{tabular} \\ \hline
\textit{Participant characteristics}                      &                                                                                                                                                                                                                                                                                                                                                                                                                                                                                                                                                                               \\ \hline
\multicolumn{1}{|l|}{Personal traits}                     & \begin{tabular}[c]{@{}l@{}}- Participants with less working experience benefit more from opportunities for informal collaboration and \\ \hspace{0.17cm}good visibility of a meeting's content and interactions.\\ - When given the choice, participants with low remote experience may favour physical meetings when they \\ \hspace{0.17cm}expect complex topics, high cognitive load or when they are unaware of the team's collaboration dynamics.\end{tabular}                                                                                                                                  \\ \hline
\multicolumn{1}{|l|}{Personal cognitive load}             & \begin{tabular}[c]{@{}l@{}}- When meetings revolve around topics that are outside of the participant's scope of knowledge or \\ \hspace{0.17cm}experience, embodied cognition and visibility benefit information processing.\end{tabular}                                                                                                                                                                                                                                                                                                                                                    \\ \hline
\multicolumn{1}{|l|}{Participation}                       & \begin{tabular}[c]{@{}l@{}}- The format of a meeting must prioritise the needs of active participants over passive ones as they will face \\ \hspace{0.17cm}more challenges regarding collaboration.\end{tabular}                                                                                                                                                                                                                                                                                                                                                                           \\ \hline
\end{tabular}%
}
\vspace{0.1cm}
\caption{Summary of impact from meeting aspects}
\label{tab:my-table}
\vspace{-0.2cm}
\end{table*}

\feature{Participation}
Not all meeting participants are expected to participate in the same way or to the same degree. Some participants can be part of the communication hub where most of the collaboration happens, while others are satellites and passive listeners that are there to \dquote{stay on the loop} or as backup to answer questions or help with the meeting organisation~\cite{venolia_hub_satellite}. In in-person meetings, all attendees are put in the communication hub. However, virtual and hybrid meetings allow participants to be passive participants without disrupting the meeting or distracting others and allowing them to work on other tasks. Within hybrid meetings, the physical space remains the primary setting for the hub participants~\cite{karis_remote}.

\underline{Impact:}
For meeting attendees who are not expected to actively participate, virtual and hybrid meetings will allow them to use their time more effectively while still being part of the discussion. The focus must then be set on the hub participants who may benefit more from the physical space depending on the values of the other model dimensions.

\section{Preliminary evaluation} 
  
 \subsection{Methodology}
 To receive external input for our model and evaluate its completeness, we utilised our professional network to reach out to experienced professionals in managerial positions who organise and conduct meetings on a regular basis and who had worked in co-located, remote and hybrid environments. We chose managers as they organise meetings more often than people in other roles and tend to have more leverage to define how and where a meeting happens. We did not discriminate for any given industry or location. In total, we had 11~participants (5~female and 6~male), all from different companies and spawning 6~different industries including healthcare~(1), banking~(3), trade and development~(2), IT services~(1), IT consulting~(3) as well as business consulting~(1). The participants' age ranged from 31 to 35~(6), 36 to 40~(2), 41 to 45~(2) and 56 to~60~(1).

We conducted individual semi-structured interviews to understand how this group of professionals made day-to-day decisions regarding meeting planning. The interview consisted of four sections where we covered their professional background, meeting culture at their companies, their meeting planning process and their opinions about recent research from meeting and collaboration topics. As an outcome of these interviews, we identified four factors that we had originally missed and which are further discussed in the following section.    

\subsection{Extensions of the Model}
\label{sec:adaptation}
We used our interview participants' insights to improve our understanding about the completeness of each dimension and added factors that were not yet addressed.

\head{Impact}
We added the impact feature to the \emph{meeting variables} dimension. It describes whether the meeting goes beyond day-to-day problem solving and requires participants to make decisions and reach agreements, referred to as \emph{strategic impact}, as well as whether the topic of the meeting is considered personal with an \emph{emotional impact}.     
When asked about what were good reasons to have an in-person meeting, most organisers answered that for meetings in which their team will make decisions with long-term impact, including vision and strategy decisions, they preferred to plan co-located in-person meetings. For instance, participant~P1 stated that \myquote{[...] it's worth it for things like planning the vision for the rest of the year [...]}. This can also be related to the fact that during strategic meetings, the ability to negotiate and persuade is very important and the capacity to do so is increased by non-verbal cues~\cite{microsoft_2021,Chidambaram_persuasive_robots}. Similarly, for topics that are considered personal, such as providing feedback, organisers were divided. For some, receiving or giving feedback remotely did not have any negative impact while for others it did. While it was difficult to express, the ones who prefer giving feedback in in-person meetings mentioned that non-verbal cues help to gauge the other person's reaction and adapt their tone and message accordingly. For sensible topics or those where privacy is required, some organisers also prefer co-located in-person meetings. We know from other Microsoft studies that meetings with a strategic or emotional impact are hard to achieve in a digital environment to the same level as for a co-located meeting \cite{microsoft_2021}.  
 
\underline{Impact:}
Strategic impact defines the need of embodied cognition, visibility and context cues. Embodied cognition will facilitate open communication and knowledge sharing as previously mentioned. The visibility of information is key to have a shared understanding of the problem to solve. Finally, context cues will help team members to remember agreements and the context in which those decisions were made for an extended period of time. Similarly, the emotional impact of a meeting influences the need for embodied cognition as emotions can increase the cognitive load of an individual~\cite{plass_four_2019}. This is especially valid for neurodivergent team members~\cite{das_towards_2021}. Embodied cognition can help to reduce that cognitive load. 

\head{Agenda and planning}
Participants brought up the importance of a clear agenda and planning to have effective and meaningful meetings. This is true for meetings across all spaces (remote, hybrid and physical)~\cite{agenda_niederman,remote_collab_msft}. Participants agreed that in meetings with a clear agenda, there is less room for non-planned discussions and \dquote{random comments}. This was perceived as a positive aspect when the team needs to stay focused but negative for ideation and meetings with creative goals. Participants also mentioned that for meetings with less well-defined agendas, they preferred to meet in the physical space where there are more spontaneous interactions and where tools can be used in more flexible ways, allowing them to improvise. These discussions are in line with previous research such as the one conducted by Microsoft~\cite{microsoft_2021}.

\underline{Impact:}
This attribute impacts the need for embodied interaction. Meetings with more flexible agendas can take advantage of embodied cognition to ease improvisation. 

\head{Trust and open communication}
In general, trust and open communication are key for collaboration~\cite{polat2018formal,laumer_not_want}. However, participants agreed that some meetings require more open communication than others, in particular those with high strategic or emotional impact. Trust and open communication are highly interconnected with embodied cognition and non-verbal cues as they are frequently used to judge personality, behaviour, intention and agreement~\cite{yang_perception_gaps}. 

One of the reasons why open communication is more seamless in person is due to the fact that people can perceive their colleagues' body language, which increases empathy. As a side effect, in-person meetings can improve relationships between team members by bringing people together in non-meeting-related activities such as going for lunch together, as mentioned by the majority of participants, or helping a colleague to move around a new building. Co-located meetings are also more prone to having unplanned interactions, increasing the chances for bonding. In general, physical interactions are better at promoting trust and team bonding. It also has a positive impact in developing the weak ties that are important for long-term innovation \cite{granovetter,microsoft_2021} 

\head{Tools}
Most tools are optimised for digital or physical spaces. For instance, a whiteboard is one of the preferred tools for collaboration but it isolates remote participants. On the other hand, digital tools are easier to be used simultaneously by co-located and remote participants but the advantages of being co-located gets lost as collaboration will mostly happen on the computing devices. Technology literacy and security compliance are two important elements why some companies still do a lot of co-located work. In the case of technology literacy, co-located collaboration normally uses everyday tools like pen and paper and hence no training is required even if participants are new to a team. Even when a specific tool is being used, other attendees can quickly help their colleagues. This is often not the case for digital collaboration, where people normally require time to be introduced to a piece of new software. The second component, security compliance, is key for industries such as banking where sensible information is managed. It implies that tools must follow stricter rules when it comes to encryption and data transfer. Even when tools comply with state-of-the-art security rules, these companies normally have lengthy approval processes for the use of any new software, meaning that teams might be stuck with old or less appropriate tools. 

\section{Informing Future Solutions}


\subsection{Guidance for Meeting Applications}
In the background section, we have highlighted the significant increase in digital meeting solutions. Major advancements are made to give users interactive tools during meetings~\cite{murali_affectivespotlight_2021}, to enhance the interaction~\cite{moira,meeting_inclusiveness}, to improve how meetings are documented~\cite{kim_learning_2014}, as well as support pre- and post-meeting flows~\cite{teevan2022microsoft}. The Cross-Space Collaboration model might be of value towards pre-meeting support. Previously, tools such as meeting schedulers have been developed with the main focus of solving issues related to time and availability constraints~\cite{purbo}. Using our Cross-Space Collaboration model, a meeting scheduler could be created that takes into account not only time constraints but attributes from the three model dimensions (e.g.~strategic impact, type of knowledge, participants' remote experience) to answer questions such as \myquote{What environment better supports the goals of the meeting? If the meeting is hybrid, who should join physically and who should join digitally? How can a meeting organiser facilitate knowledge sharing during the meeting?}. The information required by the system could be input by organisers, but the system may also keep track of variables, such as how many remote meetings participants have attended so far and automatically update their remote experience level. In other words, the dimensions of our model are the information that the system will require to define to which degree a feature (e.g.~context cues) is needed. We are aware that all teams and companies have unique requirements. With this in mind, users could customise the weight each factor has on the final recommendation to account for the unique circumstances of their situation. Such a system might be adapted so that it aligns with the team's language and day-to-day meetings. 
To better understand this use case, we envision a simple scenario where a manager is planning to schedule a meeting with their team to kickstart a project:

Maria is a project manager at a technology consulting firm. Her team of seven members is going to start a new project for a client. Maria wants to have an internal kick-off meeting to present and introduce the team to the client, to the project and its foreseeable challenges; a high strategic-impact meeting. She also wants the team to start discussing possible solutions. There is a lot of information to share, some of it is well documented on slides, some of it is a mix of notes and what Maria remembers from a previous meeting with the client. She expects the meeting to take at least one hour. In Table~\ref{tab:sample_meeting}, we can see a summary of the meeting description. 

\begin{table}
    \centering
    \begin{tabular}{|l|c|} \hline 
 \textbf{Aspects}& \textbf{Scenario}\\ \hline 
          Type of knwoledge&Mostly explicit\\ \hline 
          Intrinsic complexity&Low complexity \\ \hline 
          Agenda and planning&Not clearly defined\\ \hline 
          Duration &±1 hour\\ \hline 
          Number of participants&7\\ \hline 
          Strategic impact&Very high impact\\ \hline 
          Emotional impact&No emotional impact\\ \hline
    \end{tabular}
    \vspace{0.2cm}
    \caption{Summary of the sample meeting characteristics, not including individual characteristics of participants}
    \label{tab:sample_meeting}
\end{table}

Maria inputs this information into the company's meeting scheduling system and the system considers all of the following. 
\begin{itemize}
    \item Because most of the knowledge that is going to be shared is \textbf{explicit} (i.e.~it is concrete and parts of it are documented) and because it is \textbf{intrinsically not complex}, the team's cognitive load will be lower; this means they will be able to process new knowledge efficiently even with reduced contextual cues, visibility and opportunities of embodied interaction. Thus, the meeting can be either physical, hybrid or digital. 
    \item Because the agenda is \textbf{not clearly defined}, Maria may need to improvise more in terms of how to share and explain data. She will also struggle more to moderate discussions between participants as there are no clear rules on how to participate. Both of these situations will be easier to handle in a physical meeting thanks to physical cues that facilitate interaction and major visibility of what the team is doing (e.g.~drawing a diagram).
    \item Because the meeting will most likely take \textbf{longer than 45~minutes}, a physical meeting will help participants to get less tired and increase their engagement. 
    \item Because the meeting is of \textbf{very high strategic impact}, a physical meeting will help participants to remember agreements for a longer time thanks to contextual cues and increase the chances of unplanned interactions for the team to build trust and an open communication. 
\end{itemize}

At a glance, we might think that a physical meeting is the most obvious response. However, the system takes into account the unique situation of every team, individual as well as the company's policies; for instance:

\begin{itemize}
    \item The company allows teams to work fully remotely.
    \item Two team members are low-experience employees.
    \item Two experienced team members will mostly supervise and support other teammates but not directly contribute to the project and they prefer working remotely. 
\end{itemize}

After computing all variables, the system outputs three options:
\begin{enumerate}
    \item An in-person meeting as long as the meeting requires at least one hour. A shorter meeting may be perceived as a high-effort, low-impact activity. 
    \item A hybrid meeting where the two most experienced members can join either physically or digitally to address their preferences. The system recommends digital tools to use to handle collaborative work.
    \item A fully digital meeting. This will require a clearer agenda to reduce friction during collaboration and is not recommended to extend over more than one hour. As it requires more planning, it might not align with the urgency of holding the meeting. 
\end{enumerate}

If the characteristics of the meeting were different, the results could change drastically. Imagine the cognitive load was high due to the high intrinsic complexity of the topic and the team was mostly formed by junior team members. A fully digital meeting would not be recommended by the system. On the other hand, if the meeting was for a highly experienced team that has worked together before and the project manager had very well-documented information, the system might show that the advantages of a physical meeting over a digital one are minimal. 

We have kept this example shallow in comparison to the depth of the model, but we are confident that it illustrates the possibilities of its use. 

\subsection{Lessons for Virtual and Augmented Reality}
There is growing interest from larger companies and academia to use virtual reality~(VR) as well as augmented reality~(AR) for collaboration in various scenarios, such as building construction~\cite{zaker_virtual_2018} and assisting surgeons~\cite{artemis}. When developing these types of tools, designers can use our model to point to gaps in their design or to identify areas where they can focus to maximise their efforts. For example, our model shows that context cues have a greater impact than embodied cognition in meetings where explicit knowledge is shared. This means that designers will greatly benefit from focusing on bringing unique objects and setups that directly relate to the topic being discussed and allowing users to come back to these \emph{ad-hoc} scenarios to take advantage of the reinstatement effect~\cite{essoe_enhancing_2022,shin_context-dependent_2021}. 

\subsection{Application in Education and Learning}
Adaptive learning systems could also benefit from our model. These types of systems consider unique characteristics of students, such as skills and proficiency level, to adapt how and what content they deliver to students~\cite{review_adaptive_learning,tang_reinforcement_2019}. Our model can further enhance learning environments by providing insights about which spaces and tools are best suited for specific educational activities, considering factors such as the topic’s complexity, the dynamics of the class (e.g.~collaborative vs.~individual work) as well as student and group characteristics. By taking into account the different cognitive load that a topic imposes on students, our model can indicate which ones will benefit more from a physical learning environment and which ones might choose between physical and digital. Moreover, that input can be utilised to recommend the use of specific technologies such as virtual or augmented reality which can, for instance, adapt the modality and types of context cues based on students' needs.

\section{Conclusion and Future Work}
In the last two years, collaboration and meeting culture has changed changed dramatically. Hybrid environments and cross-space collaboration have become more common, raising the question of what is the best way to meet. In addition, more companies are investing in technologies like virtual reality and artificial intelligence in an effort to advance the field of collaboration. To maximise the output of these efforts, we formalised existing knowledge in the presented Cross-Space Collaboration model. Our model bridges the gap between numerous research studies and findings in the areas of cognitive science, knowledge transfer, interaction and collaboration. It further provides valuable insights into the role that cognitive, meeting and participant aspects have in facilitating collaboration.

Nevertheless, there are several ways how future research can further enhance our findings, including synchronous and asynchronous communication and multi-directional interaction. While future work should aim to expand on these topics, we also plan to conduct a larger and more detailed study covering multiple industries and regions to further validate the completeness of our model and unveil factors that might be unique to specific industries and regions. Finally, we plan to realise a meeting scheduler based on the Cross-Space Collaboration model and conduct a long-term study of the meeting scheduler in industry settings to assess the impact of our model over a longer time period.

\balance

\bibliographystyle{ACM-Reference-Format}
\bibliography{TR2023}

\appendix

\end{document}